\documentstyle[12pt,cite,epsfig,wrapfig,subfigure]{article}        
\begin{document}
\def\lapproxeq{\lower .7ex\hbox{$\;\stackrel{\textstyle
<}{\sim}\;$}}
\def\gapproxeq{\lower .7ex\hbox{$\;\stackrel{\textstyle
>}{\sim}\;$}}
\def\etal{\it et al., \rm}
\newcommand{\fl}{\hspace*{-\mathindent}}

\newcommand{\beq}  {\begin{equation}}
\newcommand{\eeq}  {\end{equation}}
\newcommand{\bmath}{\begin{eqnarray}}
\newcommand{\emath}{\end{eqnarray}}
\newcommand{\rme}  {\rm e}
\newcommand{\rmme}  {\rm \mu e}

\renewcommand{\arraystretch}{1.2}

{\begin{center} POLARISED DEEP INELASTIC SCATTERING: \\
DATA ANALYSIS AND RESULTS\\
\end{center}}
\vskip0.5cm

\begin{center} 
Barbara Bade{\l}ek \\ 
\vskip0.5cm
Physics Institute, University of Uppsala, S-751 21 Uppsala, Sweden \\
and \\
Institute of Experimental Physics, Warsaw University, \\
PL-00 681 Warsaw, Poland \\
\end{center}
\vskip1cm

\begin{footnotesize}
Spin properties of the nucleon are discussed based on the ongoing and
planned measurements. R\^ole, method and features of radiative corrections
applied in the analyses are presented. Future prospects of the spin physics
are reviewed.
\end{footnotesize}

\setcounter{equation}{0}
\setcounter{figure}{0}
\renewcommand{\thetable}{1.\arabic{table}}

\begin{center}
{\bf 1. Introduction}
\end{center}
Interest in spin phenomena in deep inelastic scattering
revived in the eighties after the European Muon Collaboration, EMC,
 discovered \cite{emc} that the quark contribution to the proton spin 
is substantially smaller than expected. The problem of
origin of the proton spin has then led to an intense experimental and
theoretical activity. Experiments of new generation were set up, 
in which a (deep)
inelastic scattering of polarised charged leptons off polarised
proton and deuteron target was precisely studied. The experiments delivered 
very accurate and compatible data, confirming the original result of the EMC
and permitting precise QCD analyses and tests of fundamental sum rules. 
The region of low $x$ turned out to be of particular interest,
in analogy to the unpolarised deep inelastic scattering. 

In spite of all this progress and effort the main question, that about the 
origin of the proton spin has not yet been answered conclusively. 
Old questions have been replaced by new ones and new goals are being set.       
In this article we review the experimental results on spin structure functions
and derived quantities and their interpretations. In accordance with
the topic of this conference we discuss in more detail a method of applying the
radiative corrections in the data analysis.

\begin{center}
{\bf 2. Formalism}
\end{center}

The deep inelastic lepton-nucleon scattering cross section is a sum
of a spin independent term $\overline{\sigma}$
and a term proportional to the lepton helicity, $h_l=\pm 1$:
\begin{equation}
        \sigma = \overline{\sigma} - \frac{1}{2}h_l  \Delta\sigma.
\label{sig}
\end{equation}
(symbols denote double diferential cross sections).
In the one photon--exchange approximation, the differential
electroproduction spin-avera\-ged cross section, $\overline{\sigma}$, is 
related to the structure
function $F_2(x,Q^2)$ and the ratio $R(x,Q^2)$ of the cross sections
for the longitudinally and transversally polarised virtual photons by
\begin{eqnarray}
{d^2\overline{\sigma} (x,Q^2)\over dQ^2dx}  =  \nonumber \\
 = {4\pi \alpha^2\over Q^4x}\left\{1-y-
{Mxy\over 2E} + \left(1 - {2m^2\over Q^2}\right)
{y^2(1+4M^2x^2/Q^2)\over 2[1+R(x,Q^2)]}\right\} F_2(x,Q^2) \nonumber \\
\hspace*{-0.5cm}
\label{1.1}     
\end{eqnarray}
where $M$ and $m$ are masses of the proton and electron (muon)
respectively, $E$ and $\nu$ are the incident lepton energy and the energy
transfer in the target rest frame, $y=\nu /E$, $x=Q^2/(2M\nu)$ and
$\alpha$ is the electromagnetic coupling constant.
Information on the function $R(x,Q^2)$, which has so far been measured 
only in fixed-target experiments, is scarce. 
On the contrary, $F_2(x,Q^2)$ is known precisely in a wide kinematic range,
see e.g. \cite{halina}. 

In the spin dependent part of Eq.(\ref{sig}), only longitudinally polarised 
leptons will be considered.
Cross section $\Delta \sigma$ gives
only a small contribution to the total deep inelastic cross section and 
in the one photon--exchange approximation it
depends on the two structure functions $g_1(x,Q^2)$ and 
$g_2(x,Q^2)$ as follows:
\begin{equation}
\label{delta_sigma}
        \Delta \sigma=\rm cos\psi\,\Delta \sigma_{\parallel}
        + \rm sin\psi\,\rm cos\phi\,\Delta \sigma_{\perp}, \nonumber \\
\nonumber
\end{equation}
where
\begin{eqnarray}
\nonumber
        \frac{{\rm d}^2\Delta\sigma_{\parallel}}{{\rm d}x{\rm d}Q^2}
      & = & \frac{16\pi\alpha^2 y}{Q^4} \Biggl[(1 -{y\over 2}
        -{\gamma^2 y^2\over 4})g_1(x,Q^2) -{\gamma^2 y\over 2} g_2(x,Q^2)
 \Biggr], \\
\nonumber
        \frac{{\rm d}^3\Delta\sigma_{T}}{{\rm d}x{\rm d}Q^2{\rm d}\phi}
      & = & -\cos \phi
           \,\frac{8\alpha^2 y}{Q^4}\,\gamma\,\sqrt{1-y-{\gamma^2y^2 
           \over 4}} \Biggl[{y \over 2} g_1(x,Q^2) + g_2(x,Q^2) \Biggr]. \\
\hspace*{-0.5cm}
\label{polaris}
\end{eqnarray}
In the above, $\psi$ denotes the angle between the lepton
and the nucleon spin and $\phi$ the angle between the scattering plane and
the spin plane; furthermore $\Delta \sigma_{\perp}= \Delta
\sigma_{T}/\cos\phi$ and $\gamma=2Mx/\sqrt{Q^2}$ is a 
kinematical factor, small within the acceptance of high energy experiments. 

The following two cross section asymmetries are usually measured 
in the experiments:
\begin{equation}
\label{A}
A_{\parallel} = 
{\Delta\sigma_{\parallel} \over 2 \overline{\sigma}}
\hspace{1cm}
{\rm and}
\hspace{1cm}
A_{\perp} = {\Delta\sigma_{\perp} \over 2 \overline{\sigma}}.
\end{equation}
These asymmetries are expressed in terms of asymmetries $A_1$
and $A_2$, often interpreted as virtual photon--nucleon asymmetries:
\begin{equation}
A_{\parallel} = D (A_1 + \eta A_2),  \hspace{1cm} A_{\perp}  =  d (A_2 -\xi
A_1),
\end{equation}
where
\begin{equation}
\label{a1a2}
A_1={g_1-\gamma^2g_2\over F_1}, ~~~~~~~~~~~A_2=\gamma{g_1+g_2\over F_1}.
\end{equation}
$D$ and $d$, often called the depolarisation factors of the virtual photon,
depend on $y$ and on $R$; factors $\eta$ and $\xi$
depend only on kinematic variables and are small in the kinematic regions
 covered by the present experiments.  
The bounds $|A_1|\leq 1$, $|A_2|\leq \sqrt R$ are satisfied. 
From the above formulae:
\begin{equation}
g_1 \approx A_1 F_1 \approx {A_{\parallel}\over D}{F_2 \over {2x(1+R)}}
\label{bottomline}
\end{equation}

Within the QPM, the spin dependent structure function $g_1$ is
given by
\begin{equation}
   g_{1}(x) = \frac{1}{2}{\displaystyle\sum_{i=1}^{n_f}} e_{i}^2
                        [ \Delta q_{i}(x) + \Delta \bar q_{i}(x) ],
\label{g1}
\end{equation}
with $\Delta q_{i}(x) = q_i^+(x) - q_i^-(x)$, where $q^{\pm}$ are the
distribution functions of quarks with spin parallel (antiparallel) to
the nucleon spin. 
Less obvious is the meaning of $g_2$
which contains a leading twist part, completely determined by $g_1$ and a
higher twist part, the meaning of which is subject to debate~\cite{jaf_g2}.

In QCD, $g_1$ evolves according to Altarelli--Parisi
equations, similar to the unpolarised ones. Corresponding coefficient-
and splitting functions have recently been calculated 
in the $\overline {MS}$ renormalisation scheme, up to order 
$\alpha_S^2$ \cite{nlocorr}, permitting the next-to-leading order QCD
analysis of $g_1$ and thus a determination of the
polarised parton distributions, $\Delta q_i(x,Q^2)$.
The valence quark distributions $\Delta u_v(x,Q^2)$ and $\Delta
d_v(x,Q^2)$ can be determined with some accuracy from the data, while
the polarised sea quark and gluon distributions $\Delta
\bar{q}(x,Q^2)$ and $\Delta g(x,Q^2)$ are only loosely constrained by
the structure function measurements, see e.g.\cite{compare} for the
comparison of the leading order distributions. 

Contrary to $g_1$ and $g_2$, definite theoretical predictions exist
for the first moment of $g_1$, $\Gamma_1 = \int _0^1 g_1(x)\; 
{\rm d}x$: the Bjorken and the Ellis--Jaffe sum rules. 

\begin{center}
{\bf 3. Sum rules}
\end{center}
Several sum rules have been formulated for different combinations of
structure functions. Strict QCD predictions, valid for $Q^2\rightarrow
\infty$, exist for those  involving only
flavour nonsinglet contributions, e.g. the Bjorken sum rule. 
Experimental measurements of such sum rules provide a stringent test of 
fundamental QCD assumptions. They also in principle permit the extraction 
of the strong coupling constant, $\alpha_S$, from the data.
Due to the finite $Q^2$ of
the measurements, a predicted value of a sum rule is
usually presented in the form of a power series in $\alpha_S$, the
coefficients of which are directly calculated.

There is no strict QCD prediction for the sum rules containing the flavour
singlet contributions, e.g. the Ellis--Jaffe sum rules. The
reason is that singlet contributions contain an 
`intrinsic' $Q^2$ dependence
due to the anomalous dimension of the singlet axial vector current.
Testing them usually results in surprises which teach us a lot about the 
shortcomings of the  simple quark model.

In the experimental tests of the sum rules, a major source of systematic
errors is a limited experimental acceptance in $Q^2$ at each $x$ value.
This usually means that a sum rule is measured at a certain $Q^2_0$,
common to all points but at values of $Q^2_0$ which are
not sufficiently high to
exclude a contribution from nonperturbative effects (`higher twists').
Higher twist effects in the $Q^2$ dependence of $\Gamma_1$ will not be
considered here. They are likely to be negligible, at least at $Q^2>$1 GeV$^2$. 

All the sum rules involve integrations of observables 
over the whole 0$\leq x \leq$1 interval. 
This means that due to the limited experimental acceptance,
extrapolations from $x_{min}$ to 0 and from $x_{max}$ to 1 have to be
performed. These extrapolations are another source of systematic
uncertainties in the sum rules' tests. 
In particular, evaluation of $\Gamma_1$ requires extrapolations of $g_1$
to $x$ equal 0 and 1. The latter is not
critical since $g_1\rightarrow$0 at $x\rightarrow$1 but the former is a
considerable problem since $g_1$ is probably not constant as $x$ decreses
and thus its contribution to $\Gamma_1$ at low $x$ may be sizable.

\begin{center}
{\it 3.1 Low $x$ behaviour of $g_1$}
\end{center}

The data suggest a difference in the small $x$ behaviour of $g_1^p$ and $g_1^n$ 
(cf. Section 5) and that indicates a sizable non-singlet contribution
to $g_1$ in that region. 
Expectations concerning the $g_1$ behaviour at small $x$, based on
the QCD calculations are twofold:    
(1) resummation of standard Altarelli--Parisi corrections gives, \cite{qcd,bf}:
 $g_1(x) \sim exp \left [ A \sqrt {ln(\alpha_s(Q_0^2)/\alpha_s(Q^2)) ln(1/x)
}\right ]$ for nonsinglet and singlet part of $g_1$;
(2) resummation of leading powers of ln(1/$x$) gives:
$g_1^{ns}(x)\sim 1/x^{\omega_{ns}}$, $\omega_{ns} \simeq $0.4, \cite{ryskin}
and $g_1^{s}(x)\sim 1/x^{\omega_{s}}$, $\omega_n \simeq 3 \omega_{ns} > $1,
 \cite{ryskin2} where indices `$s$' and `$ns$' refer to singlet- and non-singlet
contributions to $g_1$. Inconsistent with the above
is the Regge prediction,  that $g_1^p+g_1^n$ and $g_1^p-g_1^n$ 
behave like $x^{-\alpha}$, \cite{regge}. The lowest contributing Regge
trajectories are those of the pseudovector mesons $f_1$ (for the isosinglet
combination, $g_1^p+g_1^n$) and $a_1$ (for the isotriplet combination,
$g_1^p-g_1^n$). Their intercepts are negative and assumed to be equal:
~-0.5$< \alpha <$ 0. Finally a flavour singlet contribution to $g_1(x)$ that 
 varies as (2ln(1/$x$)-1) was obtained from a model where an exchange of two
nonperturbative gluons is assumed, \cite{bass}. Even very divergent forms 
like $g_1(x) \sim (xln^2x)^{-1}$ have been considered, \cite{close}.

Results on
$\Gamma_1$ thus depend on the assumptions made in the $x\rightarrow$0 
extrapolation. Both SMC and SLAC experiments assume the Regge like 
behaviour of $g_1$, $g_1\sim x^{-\alpha}$,
with $\alpha$=0. A value of $g_1$ is evaluated as an average at the two 
lowest $x$ data points
and a resulting contribution from the unmeasured region in (low) $x$ 
to the $\Gamma_1$ is estimated. This contribution is then taken as
a measure of the corresponding contribution to the systematic error 
on $\Gamma_1$.
 
\begin{center}
{\it 3.2. The flavour nonsinglet (Bjorken) sum rule}
\end{center}

This sum rule was obtained by Bjorken \cite{bjorken}
 from the current algebra and isospin symmetry between the proton 
and the neutron:
\begin{equation}
\label{bj_sr}
\Gamma_1^{\rm p} - \Gamma_1^{\rm n} =
\frac{1}{6} \left | \frac{g_A}{g_V} \right | = {1\over 6}(\Delta u - \Delta d)
\label{bjorken}
\end{equation}
where $g_A$ and $g_V$ are the axial and vector weak coupling constants
in the neutron beta decay and $\Delta q$ denote first moments
of the spin dependent parton distributions in the proton,
$\Delta q=\int_0^1\Delta q_i(x){\rm d}x$. This sum rule has later been
derived in the QCD and is one of the strict predictions made by this
theory.
The QCD corrections to (\ref{bjorken}) have been computed 
 up to the order $\alpha_S^3$ \cite{19} and the 
${\cal O}(\alpha_S^4)$ have been estimated \cite{20}.

\begin{center} 
{\it 3.3. The flavour singlet (Ellis--Jaffe) sum rules}
\end{center}

Separate sum rules, obtained by Ellis and Jaffe \cite{ellisjaffe},
hold for the proton and the neutron:
\begin{equation}
\label{ej_sr}
\Gamma_1^{p(n)} = \pm {1\over 12}\left | \frac{g_A}{g_V} \right | + 
{1\over 36}a_8 + {1\over 9} \Delta \Sigma
\end{equation}
Here $\Delta \Sigma = \Delta u+\Delta d+\Delta s$ is the flavour 
singlet axial coupling, $a_8=3F-D$ and 
$\left | {g_A}/{g_V} \right | = F+D$ are related to the symmetric and
antisymmetric weak flavour-SU(3) couplings ($F$ and $D$) in the baryon octet
and $\Delta q$ were defined in Sec.~3.2. If the
flavour-SU(3) is exact then $a_8$ can be predicted from measurements 
of hyperon
dacays. There is however no prediction for $\Delta \Sigma$, except when 
$\Delta s$=0. In this case $\Delta \Sigma = a_8 \sim$ 0.6, as was assumed in the
original formulation by Ellis and Jaffe~\cite{ellisjaffe}. 
QCD corrections to these sum rules have been calculated up to the order
$\alpha_S^2$ \cite{21} and the ${\cal O}(\alpha_S 
^3)$ have been estimated \cite{22}. Due to the axial anomaly of the
singlet axial vector current, $\Delta \Sigma$ is intrinsically
$Q^2$--dependent. Depending on the factorization scheme applied
\cite{lamli} this results either in a scale--dependence of the sea
quark polarization or in an extra contribution to the Ellis--Jaffe sum
rule,  involving $\Delta g=\int_0^1[g^+(x)-g^-(x)]\; {\rm d}x$,
the gluonic equivalent of the quark distribution moments. Both
formulations are equivalent.





\begin{center}
{\bf 4. Experiments and elements of data analysis}
\end{center}

Until recently the experimental knowledge on the spin structure
functions came entirely
from conventional fixed-target setups: EMC and Spin Muon Collaboration,
SMC, at CERN and experiments at SLAC. Now it is being complemented
by the results from the unconventional though {\it par excellance} 
fixed-target, HERMES experiment, at the HERA $e-p$ collider. 
Experiments wih polarised beams at colliders are also planned.  

The fixed-target electron (muon) scattering experiments are 
inclusive, i.e. information on the kinematic variables comes only 
from measurements of the incident and scattered leptons. Hadrons resulting
from the target breakup are also measured, however their 
identification until now was incomplete.

New generation polarised deep inelastic scattering (DIS) 
experiments are listed in Table 1 and their 
kinematic coverage is shown in Fig.1.

{
\footnotesize

\begin{table}[htb]
\begin{center}
{Table 1. New generation experiments on polarised deep 
inelastic charged lepton -- nucleon scattering. The last column 
shows references to the principal physics results obtained until 
now, (from \protect\cite{voss}, updated).}

\begin{tabular}{|l|c|c|c|c|c|}
\hline \hline
 Experiment & Beam & Year & Beam energy (GeV) & Target & References\\ \hline \hline
SMC & $\mu^+$& 1992--5 & 100,190  & C$_4$D$_9$OD & \cite{smcd93,smcd95,smcd96} \\
& & 1993 & 190 & C$_4$H$_9$OH & \cite{smcp96,smcp94,smcg2} \\
& & 1996 & 190 & NH$_3$ & \\ \hline
E142 & $e^-$ & 1992 & 19.4 --25.5 & $^3$He & \cite{e142,e142_new} \\
E143 & $e^-$ & 1993 & 29.1 & NH$_3$, ND$_3$ & \cite{e143p,e143d,e143g2} \\
E154 & $e^-$ & 1995 & 50 & $^3$He & \cite{e154} \\
E155 & $e^-$ & 1996 & 50 & NH$_3$, ND$_3$ & \\ \hline
HERMES & $e^-$ & 1995-- & 30--35 & H, D, $^3$He & \cite{hermes} \\
\hline \hline
\end{tabular}
\end{center}
\label{table7.1}
\end{table}

}

In DIS experiments the low $x$ region is correlated
with low values of $Q^2$ and the range of $Q^2$ covered at low $x$ is very
limited. The lowest values of $x$ were reached by the SMC at CERN by
applying special experimental techniques permitting
measurements of muon scattering angles as low as 1 mrad. These `small $x$
triggers' and special off-line selection methods were also effective against
the background of muons scattered
elastically from target atomic electrons which produce a peak at
$x=$0.000545.  

\begin{figure}[ht]
\begin{center}
\epsfig{figure=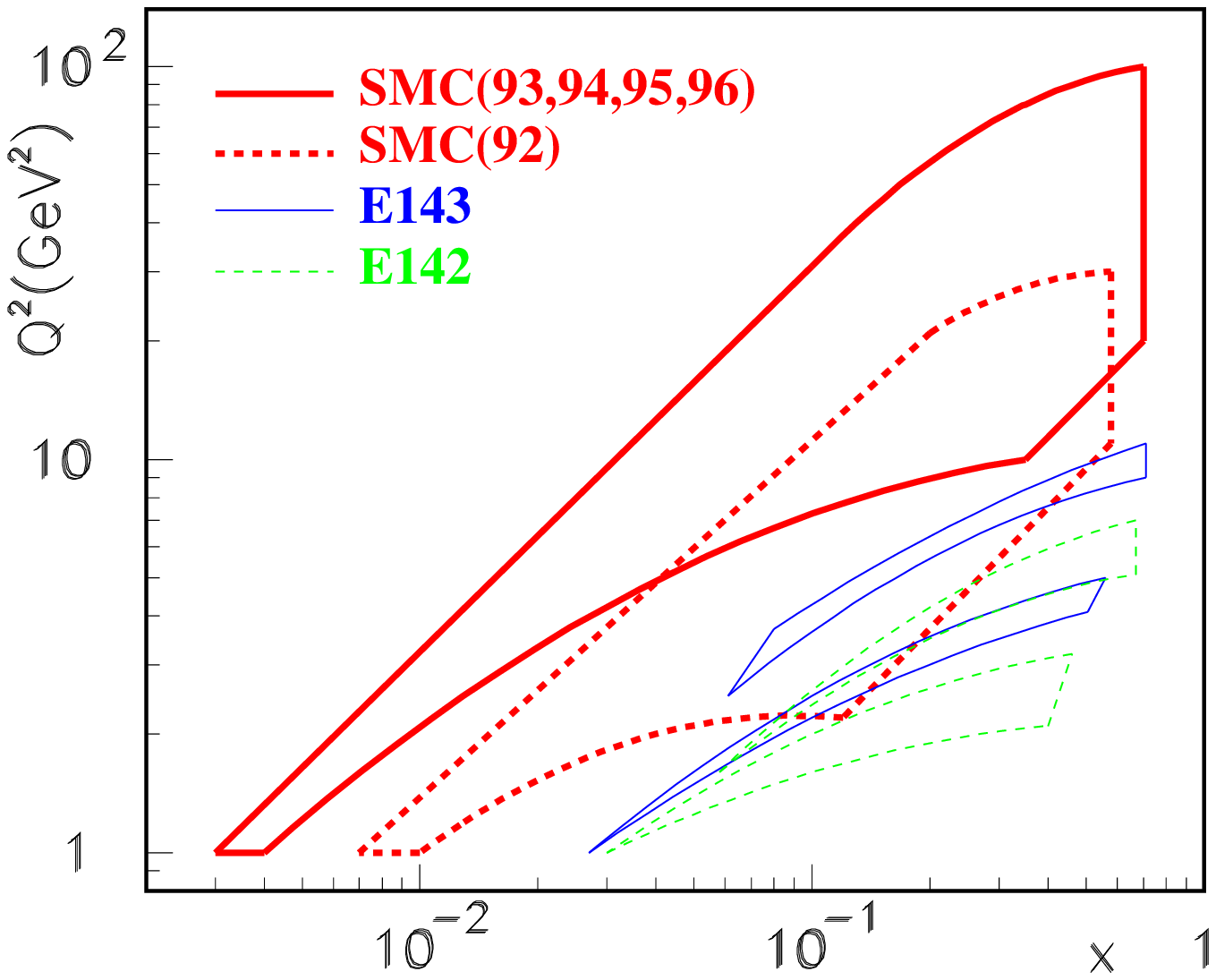,width=10cm}
\end{center}
{\footnotesize Figure 1. 
 Kinematic range of measurements by certain polarised
DIS experiments. Two areas for SMC refer to runs with 100 (1992) and
190 (1993--1996) GeV incident muon energy. For each of the two SLAC
experiments, E142 and E143, two areas correspond to two spectrometer
arms. The SMC has recently extended their analysis to
$x\sim$10$^{-5}$ and $Q^2$ values substantially lower than 1 GeV$^2$.
Acceptance of HERMES slightely extends that of SLAC experiments.
 Figure taken from \protect\cite{akio}.
}
\label{acceptance}
\end{figure}

Charged lepton DIS experiments benefit from high rates and
low (unfortunately complicated) systematic biases. They however have to deal
with a strong $Q^2$ dependence of the cross section (photon propagator 
effects) and with large contribution of radiative processes.
Electron and muon measurements are complementary: the former offers
very high beam intensities and thus statistics but its kinematic
acceptance is limited to low values of $Q^2$ and moderate values
of $x$, the latter extends to higher $Q^2$ and
down to low values of $x$ (an important aspect in the study of sum rules) but
due to limited muon intensities the data
taking time has to be long to ensure a satisfactory statistics.

The SMC experiment  
at CERN uses a naturally polarised muon beam ($\sim$80 $\%$ polarisation) 
and a double-cell, cryogenic, solid state target. 
The beam polarisation at the SMC has been measured
with a purpose-built polarimeter, using two independent methods:
polarised $\mu e$ scattering and an analysis of the energy spectrum of
electrons coming from the muon decay.
Average polarisation of the target was 
about 86$\%$ and 50$\%$ for the butanol (proton) and deuterated butanol
(deuteron) respectively. Experiments E142 -- E155 at
SLAC use an electron beam (polarisation about 86$\%$; E142 - 40$\%$)
and liquid (solid) cryogenic targets (polarisation 
 reached 80$\%$ for the proton and 25$\%$ for the deuteron one in
E143; for the $^3$He gas target: $\sim$30$\%$). The
HERMES experiment at DESY uses a self-polarised (in $\sim$50$\%$) 
electron beam from the HERA collider
and internal gas targets (polarisation $\sim$50$\%$ for $^3$He). 
Frequent exchange of target- (SMC, HERMES) and beam (SLAC) polarisations
permitted to greatly reduce systematic errors on cross section asymmetries.
The scattered muon spectrometers in the SMC and
SLAC experiments have been used (with little change) in DIS experiments
preceeding the polarised programme, contrary to the HERMES,
purpose-built apparatus. 

The cross section asymmetry measured in the polarised lepton -- polarised
nucleon experiments, $A_{exp}$, is related to the
asymmetries defined in eq.(\ref{A}) by 
\begin{equation}
A_{exp}= fP_tP_bA
\label{aexp}
\end{equation}
where $P_t,P_b$
denote the target and beam polarisations and $f$, the target dilution
factor, accounts for the fact that only a fraction of
nucleons is polarised. Dilution factors are about 0.10-0.2 in the
SMC and SLAC and 1 at HERMES proton target (0.3 for the $^3$He target).

\begin{center}
{\it Radiative corrections in the data analysis}
\end{center}

The structure functions, polarised as well as unpolarised were defined 
for the one-photon exchange reaction, cf. equations 
(\ref{1.1}),(\ref{polaris}). Higher order QED corrections,
which we have ignored so far, have thus to be applied to the measured
asymmetries, (\ref{aexp}), to convert them to the single-photon
asymmetries. These `radiative corrections' have to be applied in two places:
in the evaluation of the dilution factor  
and in the asymmetry, \cite{smc_dil_note}. 
Below we give a short account of the method used by the SMC; methods
used by HERMES and SLAC are similar.

%

\begin{figure}[htb]
\begin{center}
\setlength{\unitlength}{0.1mm}
\begin{picture}(1200,650)
\put( 00, -750){
\epsfig{file=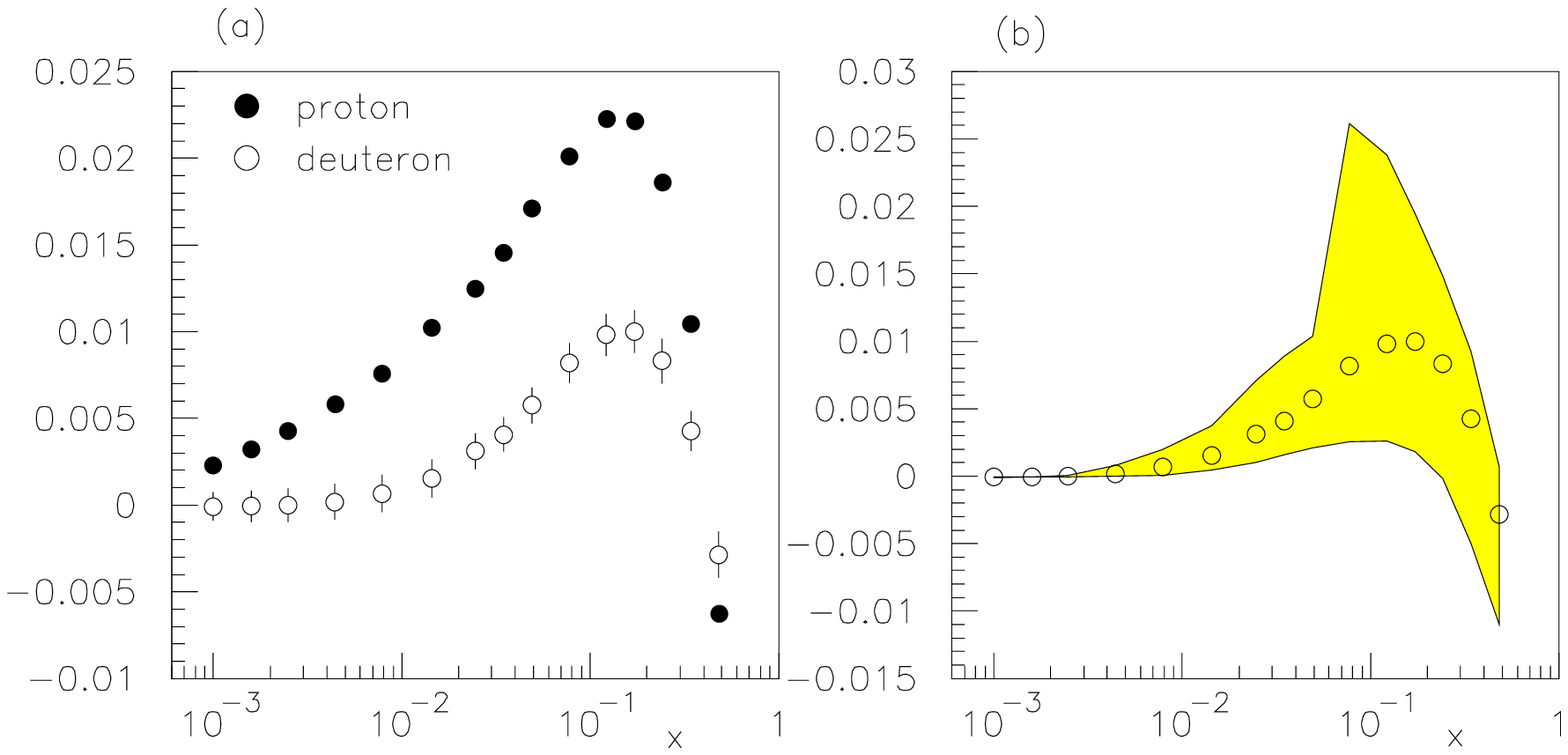,width=130mm,height=150mm}
}
\end{picture}
\end{center}
{\footnotesize Figure 2.
(a) Average radiative correction term $(\delta A_1)^{rc}$
for the proton and deuteron target as a function of $x$.
(b) Same term for the deuteron target. 
A band around points shows a variation of the
correction with $Q^2$ (or $y$) in each bin of $x$.
Figure taken from \protect\cite{akio}.
}
\label{rc}
\end{figure}

Understanding of the radiative corrections procedure will be faciliated by
introducing in this section an extended notation. The measured
observables will acquire superscripts `t' (`total`, i.e. comprising all
radiative processes) and one-gamma exchange functions -- superscripts
`1$\gamma$'. In this way, a cross section measured in a polarised experiment
is related in the following way to the one-gamma cross section:
$\sigma^t = v\sigma^{1\gamma} + \sigma_{tail}$. Analogous relation holds for
spin-independent cross section:
$\overline{\sigma}^t=v\overline{\sigma}^{1\gamma}+\overline{\sigma}_{tail}$.
Here $v$ which mostly accounts for vacuum polarization,
was found to be close to unity and thus subsequently put equal to 1;
$\sigma_{tail}$ ($\overline{\sigma}_{tail}$) are contributions
from the elastic, quasi-elastic
and inelastic lepton--nucleon and lepton--nucleus scattering.
A direct consequence of the above equations is the following relation between
a measured- and one-photon exchange asymmetries:
\beq
A_1^t=\rho [ A_1^{1\gamma}+ (\delta A_1)^{rc} ].
\label{corr_asy}
\eeq

The factor $\rho=v\overline{\sigma}^{1\gamma}/\overline{\sigma}^t$
in the above equation was eveluated using the program TERAD~\cite{terad}
and its value differed from unity at most by 2$\%$.
The additive term $(\delta A_1)^{rc}=(\delta A_{\parallel} /D)^{rc}=
(\Delta{\sigma})_{tail}/2Dv\overline{\sigma}^{1\gamma}$, was evaluated
using the program POLRAD~\cite{shumeiko,shumeiko2}.
The factor $\rho$ 
has been incorporated in the evaluation of the dilution factor, see below.      
Magnitude of the additive correction $(\delta A_1)^{rc}$ is displayed in
Fig.2.    
 
The additive correction was evaluated for both $A_{\parallel}^t$ and $A_{\perp}^t$.
The asymmetry $A_1^{\rm p}(x)$ required for these calculations in POLRAD
is taken from Refs.~\cite{emc,smcp94,e143p}. In the longitudinal case
the contribution from $A_2^{\rm p}$ is neglected.
The uncertainty of $(\delta A_1)^{rc}$ is estimated by varying the
input values of $A_1^{\rm p}$ within the errors.
The radiative corrections to the transverse asymmetry $A_{\perp}^t$ were
evaluated assuming that $g_2 = g_2^{\rm WW}$~\cite{gww}.
The corrections are much smaller than the statistical error of $A_{\perp}$ and
therefore the additive correction has been neglected.
 
%
%
%
In addition to butanol (or deuterated butanol), the target cells contain
other chemical elements. Thus
the dilution factor $f$ can be expressed in terms of the number  $n_A$  of
nuclei with mass number $A$ and 
the corresponding spin-averaged cross sections  
$\overline{\sigma}_A$ per nucleon, which include   higher order
QED effects,  for all the elements involved:
\begin{equation}
\label{dil}
f = \frac{n_{\rm H}\cdot\overline{\sigma}_{\rm H}^t}
              {\sum_A n_A\cdot\overline{\sigma}_A^t}.
\end{equation}
The one-photon exchange cross-section ratios
$\overline{\sigma}_A^{1\gamma} / \overline{\sigma}_{\rm H}^{1\gamma}$
for D, He, C and Ca
required for the calculation of $f$
are obtained from the
structure function ratios $F_2^{\rm d}/F_2^{\rm p}$~\cite{NMCR}
and $F_2^A/F_2^{\rm d}$~\cite{nmc_adep}.
The cross section ratios
$\overline{\sigma}_A^{1\gamma} / \overline{\sigma}_{\rm H}^{1\gamma}$
are converted to
$\overline{\sigma}_A^t/\overline{\sigma}_H^t$
using TERAD.
For unmeasured nuclei the cross section ratios are obtained
in the same way from a
parameterization of $F_2^A(x)/F_2^{\rm d}(x)$ as a function of $A$, 
\cite{aparam}.

For the actual evaluation of asymmetries (\ref{A}) employing  (\ref{aexp}) 
we use an effective dilution factor $f^\prime$ 
\begin{equation}
\label{dil_rad}
f^\prime = \rho f,
\end{equation}
which accounts for the multiplicative part of the radiative correction to the
asymmetry by including $\rho$ as part of an event weight.

The  dilution factors $f$ and $f^\prime$ for the proton target
are shown as a solid and broken lines in Fig.~3
and are compared to the `naive' expectation for a mixture of
62\% butanol, ($\rm CH_3(CH_2)_3OH$),
and 38\% helium by volume, i.e., $f \simeq 0.123$.
The rise of $f$ at $x>0.3$ is due to the strong decrease with $x$ of the ratio
$F_2^{\rm d}/F_2^{\rm p}$, whereas the drop in the low $x$-range
is due to the dilution by radiative events.
%
%

\begin{figure}[ht]
\vspace*{-1cm}
\begin{center}
\epsfig{figure=fig9.ai,width=7cm}
\end{center}
\vspace*{-1.cm}
{\footnotesize Figure 3. The dilution factor  $f$
as a function of $x$ for the SMC proton target (solid line).
The dashed line shows an effective dilution factor, $f\protect{^\prime}$.
The horizontal dashed line shows the naive expectation.
Figure taken from \protect\cite{smcp96}.

}
\label{fig3}
\end{figure}

Radiative events populate heavily the lowest $x$ bins of the observables
in the SMC kinematic range. These events are corrected for only on the level 
of the asymmetry determination. However they constitute a background and thus
should be removed from the sample 
before the statistical accuracy of measurements is determined.
The way the SMC applied radiative corrections in their previous analyses resulted 
in retainig these events for the statistical error determination and, 
as a consequence, in the underestimation 
of statistical errors on asymmetries, especially at low $x$. 
The new procedure, described above, guarantees a proper calculation of the
statistical error in the asymmetry, in  contrast to the previous SMC
analyses~\cite{smcp94,smcg2,smcd93,smcd95} where 
the formula $A_1^t = A_1^{1\gamma} + (\delta A_1)^{rc}_{old}$ was
used instead of Eq.(\ref{corr_asy}).
The new  procedure resulted in an increase in the $A_1$ statistical
error by a factor of $1/\rho$ which reaches 1.4 at smallest $x$.
It will be introduced in the forthcoming SMC publications, \cite{smcp96,smcd96}.
If should be mentioned that measured values of the asymmetries remain 
(practically) unaffected by the change in the method. 
Details of the old and new procedures are given in \cite{smc_dil_note}.
\begin{figure}[hb]
\begin{center}
\vspace*{-0.5cm}
\hspace*{-0.5cm}
\epsfig{file=fig12.ai,width=13cm}
\end{center}
\footnotesize{
Figure 4. The virtual photon asymmetry $A_1^p$ as a function of $x$ from
SMC, EMC, SLAC E80, E130 and E143. Error bars are
statistical. Figure taken from \protect\cite{smcp96}.
}
\label{fig4}
\end{figure}

\pagebreak

\begin{center}
{\bf 5. Results of the measurements and spin structure of the nucleon}
\end{center}

\begin{center}
{\it 5.1. Results for asymmetries, spin structure functions and their moments}
\end{center}

Cross section asymmetries $A_1$ and spin dependent structure 
functions $g_1$ have
been measured for the proton and deuteron targets by the
SMC, \cite{smcd93,smcd95,smcp94,smcp96,smcg2,smcd96} and by the E143, 
\cite{e143p,e143d}. 

Information on the
neutron has been evaluated from the data on $^3$He 
(E142, \cite{e142,e142_new}, E154, \cite{e154}, HERMES, \cite{hermes})
and from the data on the proton and deuteron (SMC, \cite{smcd93,smcd95,smcd96}).
All data sets are in a very good mutual agreement even if $A_1$, 
extracted from data covering different $Q^2$ intervals, has
been assumed to be $Q^2$ independent. 

Results on $A_1^p$ from different experiments are shown in Fig.4. 
The average $Q^2$ of SMC and SLAC data is different by a factor of 7 thus
suggesting that within the present accuracy, no $Q^2$ dependence of $A_1^p$ is
observed in the data -- a conclusion holding also for $A_1^d$, \cite{smcd96}
and in both cases confirmed by direct studies. The SMC 
measurements at $Q^2<$ 1 GeV$^2$, shown in Fig.4, were not used in the analysis 
of $g_1^p$ and evaluation of moments. 

\begin{figure}[hb]
\vspace*{-1cm}
\begin{center}
\epsfig{file=fig13.ai,width=9cm}
\end{center}
\footnotesize{
Figure 5. 
 Results for the $A_2^p(x)$ at $Q^2=$5 GeV$^2$. The solid line shows
the limit $|A_2|<\sqrt{R}$. Data from E143 are extrapolated to the same $Q^2$
assuming that $\sqrt{Q^2}A_2$ scales. Figure taken from \protect\cite{smcp96}.
}
\label{fig5}
\end{figure}

Results on $A_2^p$ are shown in Fig.5, \cite{smcp96}; together with the results
for $A_2^d$, \cite{smcd96}, they show that this function
is significantly smaller than the bound $\sqrt{R}$ and consistent with zero.

Conversion of $A_1$ to $g_1$, which was made under an assumption 
that $A_1$ scales, needs information on
the structure function $F_1$ or, equivalently, $F_2$ and $R$ 
and about $A_2$ (cf.
eq.(\ref{a1a2})). The NMC parametrisation of $F_2(x,Q^2)$
\cite{NMCF2} and the SLAC 
parametrisation of $R(x,Q^2)$ \cite{rworld} have been used by both
SMC and SLAC. However $g_1$ at average $Q^2$ is nearly (i.e.apart from
radiative corrections) independent of $R$ if the same $R$ is used
in extraction of $F_2$ and $g_1$. In the SMC data analysis $A_2$ was 
neglected; SLAC also assumed $A_2$=0, except in the
E143 proton analysis, \cite{e143p} where the measured $A_2$ was employed.

Results on $g_1$ for proton, deuteron and neutron $g_1$
for the SMC measurements are shown in Fig.6,
\cite{fabienne}. 
Here $g_1^n=2g_1^d/(1-1.5\omega _D) - g_1^p$ where $\omega _D 
\sim$ 0.05 is the probability of the \-D-state of the deuteron. 
A very precise, though kinematically limited ($x>$0.015) measurement of $g_1^n$ 
has been presented by the SLAC E154 Collaboration, \cite{e154}.
The behaviour of the $g_1^p$ seems to be different from that of $g_1^d$ 
and $g_1^n$,
especially at low $x$. This should be contrasted with the unpolarised 
case where a small difference between proton and neutron structure functions
can be explained by nuclear shadowing in the deuteron, \cite{newnp}.

\begin{figure}[htb]
\vspace*{5cm}
\begin{center}
\hspace*{-3cm}
\epsfig{file=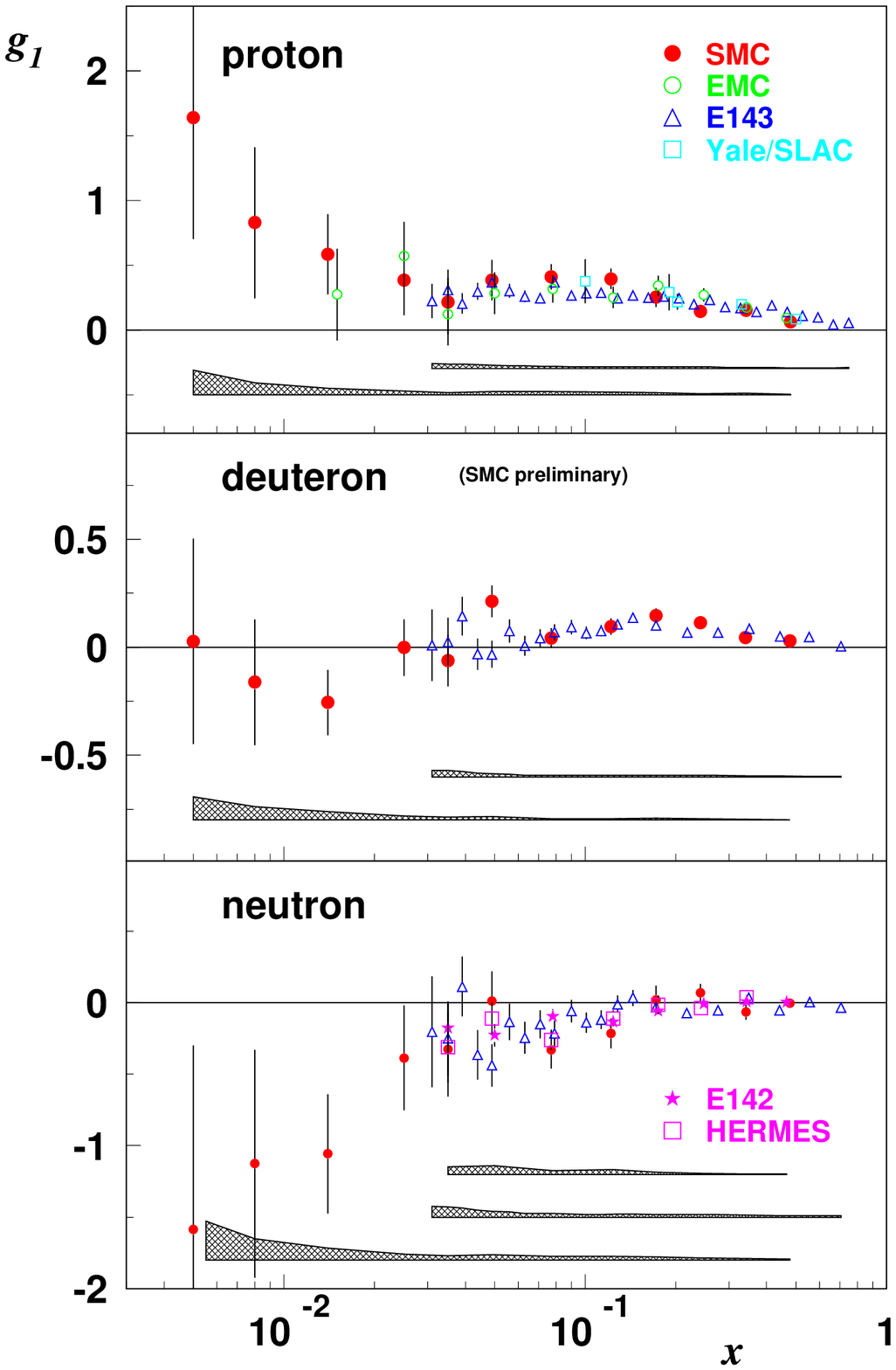,width=11cm}
\end{center}
\vspace*{-3cm}
\footnotesize{Figure 6. 
 Structure functions $g_1^p$, $g_1^d$ and $g_1^n$
at the measured $Q^2$. Error bars are statistical.
The shaded areas show the systematic errors.
Figure taken from \protect\cite{fabienne}.
}
\label{fig6}
\end{figure}

To evaluate first moments of $g_1$, a measured $g_1(x,Q^2)$
must be evolved to a scale $Q^2_0$, common for all $x$. Previously $g_1(x,Q^2_0)$
was obtained assuming $A_1 \simeq g_1/F_1$ to be independent of $Q^2$, which is
consistent with the data. However QCD predicts the $Q^2$ dependence of $g_1$ and
$F_1$ to differ considerably at small $x$ where the experimental acceptance
is very limited in $Q^2$. Therefore the data do not constrain the QCD evolution
in the region where large extrapolations in $Q^2$ are required.
Recently calculations of the NLO splitting functions
 were completed (in the $\overline {MS}$ scheme), \cite{split1,split2,split3}
thus making it possible to perform the NLO QCD evolution of the $g_1$,
\cite{evol1,evol2,evol3}.
The SMC used the
procedure \cite{evol1} to fit their proton and deuteron
data as well as these of EMC and E143. Preliminary results are
shown in Fig.7. Differences between renormalisation schemes and
values of strong coupling constants may still change the results
considerably.

\begin{figure}[htb]
\begin{center}
\vspace*{-3.cm}
\epsfig{file=fig15.ai,width=11cm}
\end{center}
\footnotesize{Figure 7. Results on $g_1^p$ and $g_1^d$ from SMC and E143 
at the measured 
$Q^2$. Solid curves are NLO QCD fits at $Q^2$ of data points, dashed -- at 
$Q_0^2$ =10 GeV$^2$ and dot-dashed at $Q^2_0$ =1 GeV$^2$.
Figure comes from \protect\cite{smcp96}.
}
\vspace*{0.5cm}
\label{fig7}
\end{figure}

\pagebreak

\begin{center}
{\it 5.2. Tests of the sum rules}
\end{center}

Table 2 shows the collected results for the first moments $\Gamma_1$ of
$g_1$ for proton, deuteron and neutron, \cite{fabienne}, assuming scaling of
$A_1$.
The SMC results for deuteron, \cite{smcd96}, and HERMES results,
\cite{hermes}, are preliminary. Numbers in parentheses are statistical 
and systematic errors respectively. The SMC and E143 neutron data result
from combining the proton and deuteron results; extrapolating
the neutron results to $x$=0 is a source of major systematic
uncertainties, especially for the SLAC data. Results for the SMC proton
change by 0.007 if instead of the $A_1$ scaling assumption, 
the NLO QCD fit is used to evolve the data to a common value of $Q^2$.
Predicted values of the Ellis-Jaffe sum rules were calculated using
the QCD corrections up to the order $\alpha_s^3$, three quark flavours, 
$\alpha_s(M_Z^2)$
= 0.117 $\pm $0.005, $|g_A/g_V|$=1.2573 $\pm$ 0.0028 
and $F/D$=0.575 $\pm$ 0.016. All data consistently violate that sum rules.

{
\footnotesize
 
\begin{table}[htb]
\begin{center}
{Table 2. results of the Ellis-Jaffe sum rule measurements. See text for details.
Theoretical predictions are given in the bottom section of the table.}
 
\begin{tabular}{|l|c|l|l|l|}
\hline \hline
 Experiment & $<Q^2>$  & Proton & Deuteron & Neutron \\
            & GeV$^2$  &        &          &         \\ \hline \hline
SMC & 10& 0.137~~(14)~~(10) & 0.038~~~(7)~~~(5) & -0.055~~~~~~~~~(24) \\ \hline
E143&  3& 0.127~~~(4)~~(10) & 0.042~~~(3)~~~(4) & -0.037~~~(8)~~(11) \\ \hline
E142&  2&                   &                   & -0.031~~~(6)~~~~(9) \\ \hline
HERMES&  3&                 &                   & -0.032~~(13)~~(17) \\ \hline
Ellis-Jaffe& 10& 0.170~~~(5)& 0.071~~~(4)       & -0.016~~~(5)       \\
sum rule   &  3& 0.164~~~(6)& 0.070~~~(4)       & -0.013~~~(5)       \\
           &  2&            &                   & -0.011~~~(5)       \\ 
\hline \hline
\end{tabular}
\end{center}
\label{table2}
\end{table}
 
}

Status of the Bjorken sum rule tests is shown in Fig.8.
Data from all experiments were evolved to $Q^2\rightarrow \infty$ for
comparison, using corrections to the order of $\alpha_s^3$ and constants
given above. All the data confirm the sum rule, predicted to give
0.2096$\pm$0.0004 at $Q^2 \rightarrow \infty$.

Recently a technique of Pad\'e approximants has been suggested for
calculating higher order corrections for the flavour nonsinglet
sum rules, \cite{pade}. As a result,
the Bjorken sum depends very strongly on $\alpha_s$ at small $Q^2$
permitting actually to extract the values of the strong coupling constant:
$\alpha_s(M_Z^2)$ = 0.117$^{+0.004}_{-0.007}\pm$ 0.002, \cite{pade},
where the first two errors are statistical and the last one is theoretical.

\begin{figure}[htb]
\vspace*{4.5cm}
\hspace*{-.5cm}
\epsfig{file=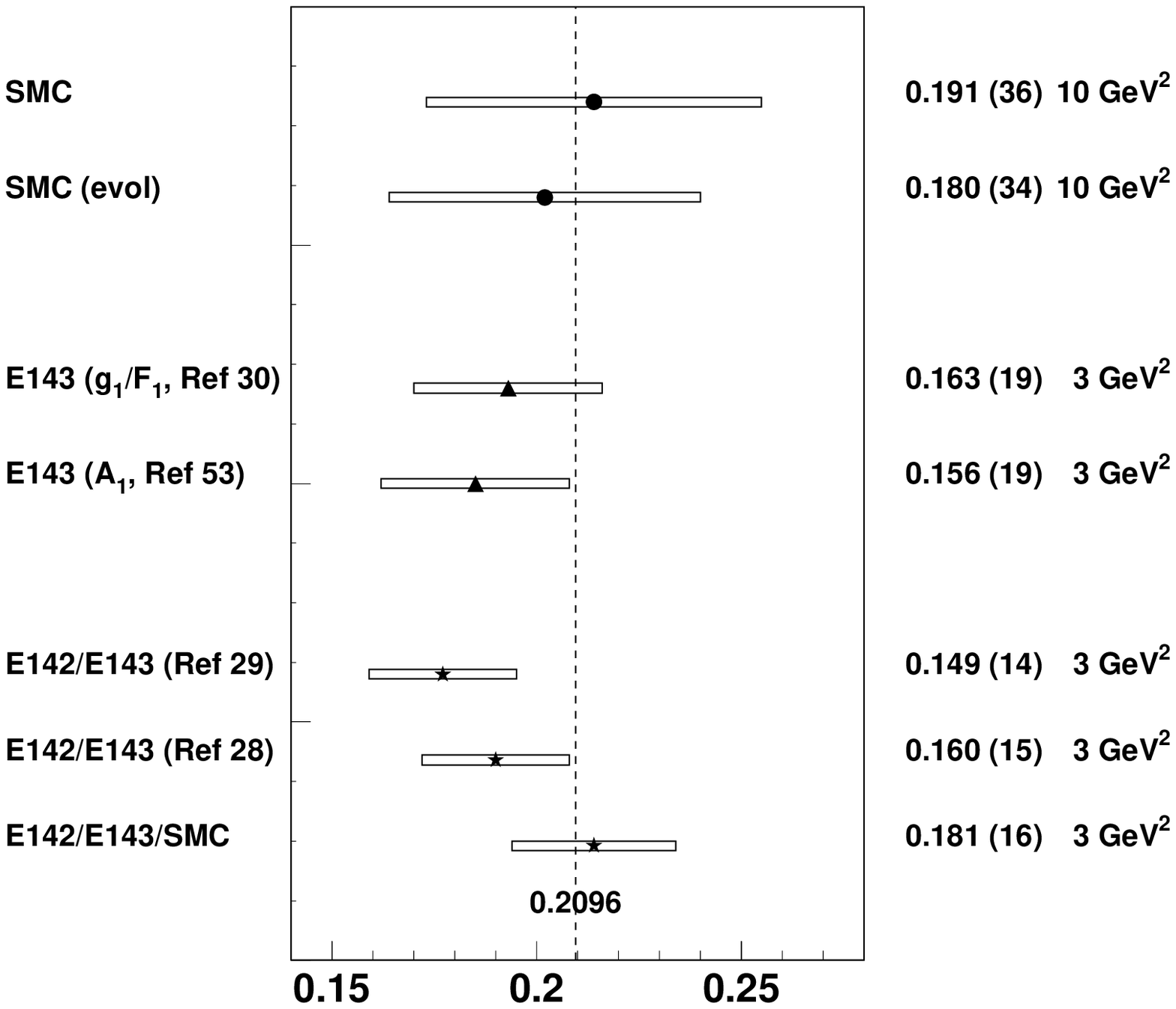,width=10cm}
\vspace*{-4cm}
\footnotesize{

Figure 8. Results on the Bjorken sum. Data are evolved to
$Q^2\rightarrow \infty$ for comparison. The SMC data are preliminary.
E143 result assuming that $A_1$ scales instead of $g_1/F_1$ was taken from
\cite{e143a1}.
Figure comes from \protect\cite{gerdhab}.
}
\label{fig8}
\end{figure}

\begin{center}
{\it 5.3. Spin structure of the proton}
\end{center}

The nucleon spin, $S_z={\textstyle\frac{1}{2}}$, can be decomposed as follows
\begin{equation}
S_z={\textstyle{1\over 2}}\Delta \Sigma + \Delta g + L_z
\label{spin}
\end{equation}
where $L_z$ is angular momentum due to the partons.
Results for $\Gamma_1$ shown in Table 2, evolved to $Q^2\rightarrow \infty$
using corrections up to $\alpha_s^3$ together with constants given
in the previous section
give the following estimate of the flavour singlet axial coupling,
 $\Delta \Sigma$ and the polarisation of the strange sea quarks, $\Delta s$.
Result is: $\Delta \Sigma \simeq$ 0.28 $\pm$ 0.07 and $\Delta s\simeq$
-0.11$ \pm$ 0.03 confirming the original EMC conclusion that quark spin
contributes little to the proton spin and that the strange sea 
is indeed polarised opposite to the nucleon spin.
The flavor SU(3) breaking (SU(3) was assumed to be exact in the 
derivation of the above numbers) can decrease 
$\Delta s$ but leaves $\Delta \Sigma $ unchanged.
Choosing a factorization scheme in which the quarks polarisation is
scale independent, a $Q^2$ dependent gluonic contribution appears in the 
Ellis-Jaffe sum rule as a result of the anomalous dimension of the singlet
axial vector current \cite{lamli}. Then the Ellis-Jaffe assumption of
$\Delta s$=0 implies that at $Q^2$=10 GeV$^2$, $\Delta g\sim 3$ is needed
to restore the sum rule. This result is compatible with conclusions
from certain QCD analyses, \cite{evol1}.

\begin{center}
{\it 5.4. Semi--inclusive results}
\end{center}

Finally we note the measurements of the semi--inclusive spin
asymmetries for positively and negatively charged hadrons in the polarised
muon--proton and muon--deuteron scattering in the SMC \cite{ww,fabienne}.
Analy\-sing the charged hadrons is the only way of separating quark flavours 
in the neutral current deep inelastic scattering.

The $x$ dependence of the spin distributions for the up and down valence quarks 
and for the non-strange sea quarks has been determined.
The up valence quarks have positive polarisation at all $x$, while down valence
ones are polarised negatively with respect to the proton spin.
The moments of the quark spin distributions
were obtained to be: $\Delta u_v$=0.85$\pm$0.0.14$\pm$0.12, 
$\Delta d_v=-$0.58$\pm$0.16$\pm$0.11 and 
$\Delta \bar q$=0.02$\pm$0.06$\pm$0.03. Here $\Delta \bar q=\Delta \bar u=
\Delta \bar d$. 

Precise semi--inclusive results are soon expected to come from the HERMES
experiment at HERA, cf \cite{hermes}.

\begin{center}
{\bf 6. Summary and prospects for the future}
\end{center}

Understanding of the polarised structure functions has improved
dramatically in the recent years, thanks to the EMC, SMC and SLAC
measuremets. Several questions however remained unanswered. Among them is
the low $x$ behaviour of $g_1$ (somewhat analogous to the
unpolarized case), its $Q^2$ evolution, the gluon
polarisation and flavour decomposition of polarised parton distribution. 
The HERMES experiment will
especially address the last question from a presently unique
reconstruction of the hadronic final state.
To answer the remaining questions
a new generation of experiments, e.g. at the HERA {\it collider}, is needed. 
Prospects of spin physics at HERA were discussed at a workshop at
DESY--Zeuthen in August 1995, \cite{herap}. 
A polarised deep inelastic programme at HERA
could allow measurements over an extended kinematic range,
including low $x$ and high $Q^2$. Polarisation of the
proton beam is technically much more complicated than polarisation
of the electron beam, as the proton beam does not polarise naturally. 
Construction of the polarised proton beams of energy up to 250 GeV
in the RHIC collider rings has already been approved, a helpful step for HERA.
Unfortunately interpretation of hadron--hadron results from the quark point
of view is certainly more complicated.
Dedicated measurements of $\Delta g(x,Q^2)$ are however crucial.
A precise result can be supplied by the COMPASS project at CERN
where `open charm' production will tag the photon--gluon fusion.
Another possibility would be to tag it through measurements of three jets
at the HERA collider, cf. \cite{fabienne}.
Naturally for the fixed-target data, the non-perturbative
effects interfere with the low $x$ dynamics. 
So there is little doubt that the spin physics will continue to be  
a field of particular interest.  

\pagebreak

\begin{center}
{\bf Acknowledgements}
\end{center}
My thanks to the organizers for the splendid conference and for
supporting my attendance and to my colleagues
from the NMC and SMC for the enjoyable research collaboration.

This research has been supported in part by the Polish State Committee
for Scientific Research grant number 2P03B 184 10.

\vskip1cm
\begin{center}
{REFERENCES}
\end{center}
\vspace*{-1.8cm}


\begin{thebibliography}{999}
%
\bibitem{emc} EMC: J. Ashman \etal Phys. Lett. {\bf B206} (1988) 364;
Nucl. Phys. {\bf B328} (1989) 1.
%
\bibitem{halina} H. Abramowicz, plenary talk on the International Conference on
High Energy Physics, Warsaw, July 1996.
%
\bibitem{jaf_g2} R.L.\ Jaffe, Comm. Nucl.\ Phys.\ {\bf 19} (1990) 239.
%
\bibitem{nlocorr} E.B. Zijlstra and W.L. van Neerven, Nucl. Phys. 
{\bf B417} (1994) 61; \\
R.\ Mertig and W.L.\ van Neerven, Leiden preprint INLO-PUB-6/95 (revised);\\
W.\ Vogelsang, Rutherford Laboratory preprint RAL-TR-95-071.
%
%
\bibitem{compare} T.\ Gehrmann and W.J.\ Stirling, Proc. of the Workshop on the
 Prospects of Spin Physics at HERA, Zeuthen, August 1995, eds
 J.\ Bl\"umlein and W.-D.\ Nowak, DESY 95-200, p.295.
%
\bibitem{qcd} M.A. Ahmed and G.G. Ross, Phys. Lett., {\bf B56} (1975) 385.
%
\bibitem{bf} R.D. Ball, S. Forte and G. Ridolfi, Phys. Lett. 
{\bf B378} (1996) 255.
%
\bibitem{ryskin} J. Bartels, B.I. Ermolaev and M.G. Ryskin, Z. Phys.
{\bf C70} (1996) 273.
%
\bibitem{ryskin2} J. Bartels, B.I. Ermolaev and M.G. Ryskin, DESY preprint
96-25 (1996).
%
%
\bibitem{regge} R.L. Heimann, Nucl. Phys. {\bf B64} (1973) 429.
\bibitem{bass} S.D. Bass and P.V. Landshoff, Phys. Lett. {\bf B336} (1994) 537.
\bibitem{close} F.E. Close and R.G. Roberts, Phys. Lett. {\bf B316} (1994) 257.
%
\bibitem{bjorken} J.D.\ Bjorken, Phys. Rev. {\bf  148} (1966) 1467; 
 {\it ibid.} {\bf D1} (1970) 465; {\it ibid.} {\bf D1} (1970) 1376.
%
\bibitem{19} S.A.\ Larin, F.V.\ Tkachev and J.A.M.\ Vermaseren, 
 Phys. Rev. Lett. {\bf 66} (1991) 862; S.A.\ Larin and J.A.M.\ Vermaseren,
 Phys. Lett. {\bf B259} (1991) 345.
%
\bibitem{20} A.L.\ Kataev and V.\ Starchenko, Mod. Phys. Lett. {\bf A10}
(1995) 235; CERN-TH-7198/94.
%
\bibitem{ellisjaffe} J.\ Ellis and R.L.\ Jaffe, Phys. Rev. {\bf D9} (1974) 1444; 
Erratum {\it ibid.} {\bf D10} (1974) 1669.
%
\bibitem{21} S.A.\ Larin, Phys. Lett. {\bf B334} (1994) 192.
\bibitem{22} A.L.\ Kataev, Phys. Rev. {\bf D50} (1994) 5469.
\bibitem{lamli} C.S.\ Lam and Bing-An\ Li, Phys. Rev. {\bf D25} (1982) 683.
\bibitem{voss} R.\ Voss, Proc. of the Workshop on the Deep Inelastic Scattering 
 and QCD, Paris, April 1995, eds J.-F.\ Laporte and Y.\ Sirois, p.77.
\bibitem{smcd93} SMC: B.\ Adeva \etal Phys. Lett. {\bf B302} (1993) 533.
%
\bibitem{smcd95} SMC: D.\ Adams \etal Phys. Lett. {\bf B357} (1995) 248.
%
\bibitem{smcd96} SMC: D.\ Adams \etal to be submitted to Phys. Lett. {\bf B}.
%
\bibitem{smcp96} SMC: D. Adams \etal submitted to Phys. Rev. {\bf D}.
%
\bibitem{smcp94} SMC: D.\ Adams \etal Phys. Lett. {\bf B329} (1994) 399;
erratum, Phys. Lett. {\bf B339} (1994) 332.
%
\bibitem{smcg2} SMC: B.\ Adeva \etal Phys. Lett. {\bf B336} (1994) 125.
%
%
\bibitem{e142} SLAC E142 Collaboration: P.L.\ Anthony \etal Phys. Rev. Lett.
{\bf 71} (1993) 959. 
%
\bibitem{e142_new} SLAC E142 Collaboration: P.L.\ Anthony \etal 
Phys. Rev. {\bf B54} (1996) 6620.
%
\bibitem{e143p} SLAC E143 Collaboration: K.\ Abe \etal Phys. Rev. Lett. 
{\bf 74} (1995) 346.
%
\bibitem{e143d} SLAC E143 Collaboration: K.\ Abe \etal Phys. Rev. Lett. 
{\bf 75} (1995) 25.
\bibitem{e143g2} SLAC E143 Collaboration: K.~Abe \etal Phys. Rev. Lett. 
76 (1996) 587.
\bibitem{e154} SLAC E154 Collaboration: E. Hughes, presented on
the International Conference on High Energy Physics, Warsaw, July 1996.
\bibitem{hermes} HERMES Collaboration: W. Wander, presented on  
the International Conference on High Energy Physics, Warsaw, July 1996.
\bibitem{akio} A. Ogawa (SMC), PhD Thesis, Nagoya University, 1996.
\bibitem{smc_dil_note} J.M. Le Goff, A. Steinmetz and R. Windmolders,
                   SMC internal note, SMC/96/09.
\bibitem{terad} A. Akhundov \etal Fortsch. Phys. {\bf 44} (1996) 373. \\
                A.A.~Akhundov \etal Sov. J. Nucl. Phys.
                { 26} (1977) 660; 44 (1986) 988;\\
                JINR-Dubna preprints E2-10147 (1976), E2-10205 (1976),
                E2-86-104 (1986);\\
                D.~Bardin and N.~Shumeiko, Sov. J .Nucl. Phys.
                { 29} (1979) 499.
\bibitem{shumeiko} T.V.~Kukhto and N.M.~Shumeiko, Nucl. Phys.
                   B219 (1983) 412.
\bibitem{shumeiko2} I.V.~Akushevich and N.M.~Shumeiko, J. Phys.
                    G20 (1994) 513.
\bibitem{gww}      S.~Wandzura and F.~Wilczek, Phys.Lett. {B72} (1977) 195.
\bibitem{NMCR} NMC: P.\ Amaudruz \etal Nucl. Phys. {\bf B371} (1992) 3.
\bibitem{nmc_adep}NMC, P.~Amaudruz et al., Z. Phys. C51 (1991) 387.
\bibitem{aparam} EMC: J. Ashman \etal Z. Phys. {\bf C57} (1993) 211.
\bibitem{NMCF2} NMC: M. Arneodo \etal Phys. Lett. {\bf B364} (1995) 107.
\bibitem{rworld} L.W.\ Whitlow \etal Phys. Lett.  {\bf B250} (1990) 193.
\bibitem{fabienne} F. Kunne-Perrot (SMC),  presented on
the International Conference on High Energy Physics, Warsaw, July 1996.
\bibitem{newnp} NMC: M. Arneodo \etal submitted to Nucl. Phys. {\bf B}
\bibitem{split1}  E.B.~Zijlstra and W.L.~van Neerven,
                   Nucl. Phys. B417 (1994) 61.
\bibitem{split2}  R.~Mertig and W.L.~van Neerven, Z.~Phys. {\bf C70} (1996) 637.
\bibitem{split3} W.~Vogelsang,  Phys. Rev. {\bf D54} (1996) 2023.
\bibitem{evol1}  R.D.~Ball, S.~Forte, and G.~Ridolfi,
                    Phys. Lett. {\bf B378} (1996) 255.
\bibitem{evol2}M.~Gluck, E.~Reya, M.~Stratmann, W.~Vogelsang,
                   Phys Rev. {\bf D53} (1996) 4775.
\bibitem{evol3}T.~Gehrmann and W.S.~Stirling,
                   Phys. Rev. {\bf D53} (1996) 6100.
\bibitem{e143a1} J. McCarthy, O.A. Rondon and T.J. Liu, Phys. Rev. {\bf D54}
(1996) 2391.
%
\bibitem{gerdhab} G. Mallot (SMC), Habilitation Thesis, Mainz University, 1996.
\bibitem{pade} J. Ellis \etal Phys. Lett. {\bf B366} (1996) 268; J. Ellis,
plenary talk at the SPIN96 conference, Amsterdam, September 1996.
\bibitem{ww} SMC: B.\ Adeva \etal Phys. Lett. {\bf B369} (1996) 93.
\bibitem{herap} Proc. of the Workshop on the Prospects of Spin Physics at HERA, 
Zeuthen, August 1995, eds J.\ Bl\"umlein and W.-D.\ Nowak, DESY 95-200, p.76-99.
%


\end{thebibliography}
\end{document}